\documentclass[12pt,a4paper]{amsart}

\usepackage{amsfonts}
\usepackage{amssymb}
\usepackage{amsthm}
\usepackage{newlfont}
\usepackage{amsmath}
\usepackage[mathscr]{eucal}
\usepackage{verbatim}
\usepackage{amscd}
\usepackage{xcolor}
\usepackage{graphicx}
\usepackage{ragged2e}

\makeatletter
\renewcommand\normalsize{\@setfontsize\normalsize{12pt}{14pt}}
\makeatother

\usepackage[colorlinks=true, linkcolor=black, citecolor=black, urlcolor=black]{hyperref} 

\newcommand{\proofbox}{$\vrule height7pt width5pt depth0pt$}

\newenvironment{pf}{\noindent \textbf{Proof}.\hspace{1em}}
                   {\hfill $\square$
                    \vspace{10pt}}

\theoremstyle{plain}
\newtheorem{lemma}{Lemma}[section]
\newtheorem{theorem}[lemma]{Theorem}

\theoremstyle{definition}

\begin{document}
\title{Monomial matrix analogue of Yoshida's theorem}

\author{Ananda Chakraborty}
\address{Department of Mathematics\\
Shahjalal University of Science and Technology\\
Sylhet-3114\\
BANGLADESH}
\email{ananda50.mat@gmail.com}

\date{\today}
\justifying
\begin{abstract}
In this paper, we study variants of weight enumerators of linear codes over~$\mathbb{F}_q$. We generalize the concept of average complete joint weight enumerators  of two linear codes over~$\mathbb{F}_q$. We also give its MacWilliams type identities. Then we establish a monomial analogue of Yoshida's theorem for this average complete joint weight enumerators. Finally, we present the generalized representation for average of $g$-fold complete joint weight enumerators for $\mathbb{F}_q$-linear codes and establish a monomial matrix analogue of Yoshida's theorem for average $g$-fold complete joint weight enumerators.

\end{abstract}



\maketitle
 \hfill
 \hfill

\bigskip
\section{Introduction}
\medskip
\justifying
 MacWilliams identities~\cite{williams} are relations that relate the weight enumerators of a linear code with that of its dual, without any information about the dual code. MacWilliams et al. ~\cite{gealson,MacWilliams} introduced the complete weight enumerator and joint weight enumerators of linear codes over~$\mathbb{F}_q$. In 1988, Yoshida~\cite{yoshida} gave the notion of average joint weight enumerators of linear codes over~$\mathbb{F}_2$. Moreover, he provided a theorem that expresses the average joint weight enumerators of~$\mathbb{F}_2$-linear codes in a significant way. We call this theorem as Yoshida's theorem. Later, Chakraborty and Miezaki~\cite{Himadri} introduced the concept of average of complete joint weight enumerators of $\mathbb{F}_q$-linear codes. Furthermore, they established its MacWilliams type identities and generalized Yoshida's theorem in~\cite{yoshida} for the average complete joint weight enumerators.

In this paper, we take average on monomially equivalent~$\mathbb{F}_q$-linear codes and generalize the concept of average complete joint weight enumerators in for two~$\mathbb{F}_q$-linear codes. Then we provide its MacWilliams type identities for it. Moreover, we establish a monomial matrix analogue of Yoshida's theorem for average joint weight enumerators of codes over~$\mathbb{F}_{q}$. Finally, we generalize the average of $g$-fold complete joint weight enumerators for linear codes over~$\mathbb{F}_q$ and present a monomial matrix analogue of Yoshida's theorem for the average~$g$-fold complete joint weight enumerators.

We organize the paper as follows. In Section~\ref{pre}, we provide basic ideas of linear codes over~$\mathbb{F}_q$ and its different weight enumerators. In Section~\ref{MacWilliamsIdentities}, we give the monomial generalize the concept of average complete weight enumerators for two~$\mathbb{F}_q$-linear codes. Then we provide MacWilliams type identities for the average complete joint weight enumerators. In Section~\ref{MAINTHEOREM}, we establish the monomial analogue of Yoshida's theorem for the average complete joint weight enumerator. Finally, in Section~\ref{Section:G-fold}, we define the average~$g$-fold complete joint weight enumerators of linear codes over~$\mathbb{F}_q$ and present the monomial matrix analogue of Yohsida's theorem for the average~$g$-fold complete joint weight enumerators.

\bigskip
\section{Preliminaries}\label{pre}
\medskip

A linear code $C$ of length~$n$ over $\mathbb{F}_q$ is a vector subspace of $\mathbb{F}_q^n$. Every element of an $\mathbb{F}_q$-linear code~$C$ is called a codeword of~$C$. Suppose $\mathbf{u}= (u_1,u_2,\ldots,u_n)$ and $\mathbf{v}= (v_1, v_2,\ldots,v_n)$ are two elements of $\mathbb{F}_q^n$. Then their inner product is defined as 
\begin{align*}
    \mathbf{u}\cdot \mathbf{v} := u_1v_1 + u_2v_2 + \cdots + u_nv_n.
\end{align*}
The dual of an $\mathbb{F}_q$-linear code~$C$ is defined as 
\begin{align*}
    C^{\perp} := \left\{\mathbf{u'} \in \mathbb{F}_q^n \mid \mathbf{u} \cdot \mathbf{u'} = 0 \,\, \text{for all } \mathbf{u} \in C \right\}.
\end{align*}
A linear code $C$ over $\mathbb{F}_q$ is called self-dual if $C = C^{\perp}$.

Let the elements of $\mathbb{F}_q$ be arranged in a particular order as $0= \omega_0, \omega_1, \ldots,\omega_{q-1}$. Also let $\mathbf{u}\in \mathbb{F}_q^n$. Then we define
\begin{align*}
   \mathrm{comp}(\textbf{u}):=s(\textbf{u})
    :=(s_a(\textbf{u})\mid a \in \mathbb{F}_q)
\end{align*} 
as the composition of $\mathbf{u}$, where $s_a(\textbf{u})= \#\{i \mid u_i = a,\, a\in \mathbb{F}_q\}$. Clearly, 
\begin{align*}
    \sum_{a \in \mathbb{F}_q^n}s_a(\mathbf{u})= n
\end{align*}
Generally, a composition $s$ of $n$ is  a vector whose components $s_a$ are non-negative integers for $a \in \mathbb{F}_q$ such that 
\begin{align*}
    \sum_{a \in \mathbb{F}_q} s_a= n
\end{align*}

Let $C$ be a~$\mathbb{F}_q$-linear code with length~$n$. Then we define 
\begin{align*}
    T_{s}^{C}:=\{\textbf{u} \in C \mid s_a=s_a(\textbf{u}),\, a \in \mathbb{F}_q\}. 
\end{align*}
 Also we denote  $A_s^{C} :=|T_s^{C}|$. Therefore, the complete weight enumerator of $C$ is defined as 
 \begin{align*}
     \mathbf{CW}_{C}(x_a \mid a \in \mathbb{F}_q)&:= \sum_{\textbf{u}\in C}\prod_{a \in \mathbb{F}_q}x_a^{s_a(\textbf{u})}\\
     &= \sum_{s}A_s^{C}\prod_{a \in \mathbb{F}_q}x_a^{s_a},
 \end{align*}
 where $x_a$'s are indeterminates.

Let $C_1$ and $C_2$ be two linear codes with length $n$ over~$\mathbb{F}_q$. 
Also let $\textbf{u}, \textbf{v} \in \mathbb{F}_q^{n}$. Then for every $(\mathbf{u}, \mathbf{v})\in C_1 \times C_2$, we define the bi-composition $\eta(\mathbf{u}, \mathbf{v})$ as
\[
\eta{(\mathbf{u}, \mathbf{v})} := (\eta_{\alpha\beta}(\mathbf{u}, \mathbf{v}) \mid \alpha, \beta \in \mathbb{F}_q),
\]
where $\eta_{\alpha \beta}(\mathbf{u}, \mathbf{v})$ are non-negative integers and
\[
\eta_{\alpha \beta}(\mathbf{u}, \mathbf{v}) := \#\{i \mid (u_i, v_i) = (\alpha, \beta)\}.
\]
 Clearly, $\sum_{\substack{\alpha, \beta \in \mathbb{F}_q}} \eta_{\alpha\beta}(\textbf{u},\textbf{v})=n$.

The complete joint weight enumerator of $C_1$ and $C_2$ is defined as 
\begin{align*}
\textbf{CJW}_{C_1,C_2}(x_{\alpha\beta}\mid \alpha, \beta \in \mathbb{F}_q) := \sum_{\substack{\textbf{u} \in C_1 \\ \textbf{v} \in C_2}} \prod_{\alpha, \beta \in \mathbb{F}_q} x_{\alpha\beta}^{\eta_{\alpha\beta}(\textbf{u}, \textbf{v})}\\
\hspace{4cm}
= \sum_{\eta} A_{\eta}^{C_1, C_2} \prod_{\alpha, \beta \in \mathbb{F}_q} x_{\alpha\beta}^{\eta_{\alpha\beta}},
\end{align*}
 where $A_{\eta}^{C_1,C_2}:= \#\{(
 \mathbf{u},\mathbf{v}) \in C_1\times C_2 \mid \eta(\textbf{u},\textbf{v})=\eta\}$.

\bigskip
\section{MacWilliams Type Identities}\label{MacWilliamsIdentities}
\medskip

 The average complete joint weight enumerators for~$\mathbb{F}_q$-linear codes was introduced in~\cite{Himadri}.The MacWilliams identities for complete weight enumerators and joint complete weight enumerators of $\mathbb{F}_q$-linear codes were presented in~\cite{williams,gealson}. In~\cite{Himadri}, the MacWilliams identities for average complete joint weight enumerators of~$\mathbb{F}_q$-linear codes were given. In this section, we generalize the average complete joint weight enumerators for two~$\mathbb{F}_q$-linear codes. Then we establish the MacWilliams type identities for~$\mathbb{F}_q$-linear codes. In the beginning, we review the concept of monomial matrix over~$\mathbb{F}_q$ and group action of the group of monomial matrices on~$\mathbb{F}_q$-linear codes.

 A square matrix $M$ with exactly one non-zero entry from~$\mathbb{F}_q^{\ast}= \mathbb{F}_q\setminus \{0\}$ in each row and column is called  a monomial matrix over $\mathbb{F}_q$. We can write a monomial matrix \(M\) as
\begin{equation*}\label{Eq:Monomial_def}
M = DP\quad \text{or}\quad M=P'D',
\end{equation*}
where \(D\), \(D'\) are invertible diagonal matrices over \(\mathbb{F}_q\) and \(P\), \(P'\) are permutation matrices. The set of all monomial matrices over $\mathbb{F}_q$ forms a group under matrix multiplication~\cite{BallState}. We denote the set of all monomial matrices over $\mathbb{F}_q$ by $\mathrm{M}_{\mathbb{F}_q}(n)$. For a~$\mathbb{F}_q$-linear code~$C$ and for any $M \in M_{\mathbb{F}_q}(n)$,  $CM=\{\mathbf{u}M \mid \mathbf{u}\in C\}$ is monomially equivalent to $C$, where $\mathbf{u}M= \{(D(P\mathbf{u}^T))^T\}$ and $\mathbf{u}^{T}$ denotes the transpose of $\mathbf{u}$. Therefore, the average complete joint weight enumerator of $C_1$ and $C_2$ over~$\mathbb{F}_q$ is defined as 
 \begin{align*}
&\mathbf{CJW^{av}}_{C_1,C_2}(x_{\alpha\beta}\mid \alpha,\beta \in \mathbb{F}_q)\\&:=\frac{1}{(q-1)^n \, n!} \sum_{M \in \mathrm{M}_{\mathbb{F}_q}(n)} \mathbf{CJW}_{C_1M,C_2}\Big(x_{\alpha\beta}\mid  \alpha,\beta  \in \mathbb{F}_q\Big),
\end{align*}

\begin{lemma}\label{Lemma:01}
     Let $C$ be a linear code with length~$n$ over~$\mathbb{F}_q$ and  $C^{\perp}$ be its dual. Let  $M$ be a monomial matrix with its inverse $M^{-1}$. Then we have 
    \begin{equation*}
        C^{\perp}M = (CM^{-1})^{\perp}
    \end{equation*}
\end{lemma}
\begin{pf} 
Let $\mathbf{y}\in (CM^{-1})^{\perp}$. Then we have
\begin{align}\label{EQU:CM_1}
\mathbf{y}\cdot\mathbf{z} = 0,
\end{align}
where $\mathbf{z}=\mathbf{u}M^{-1} \in CM^{-1}$.
Now 
\begin{align*}
\mathbf{y}\cdot \mathbf{z}&=\mathbf{y}\cdot (P^{T}(D^{-1}\mathbf{u}^T)^{T}
\\
&=\mathbf{y}\cdot((D^{-1}\mathbf{u}^T
)^TP)\\
&=\mathbf{y}\cdot (\mathbf{u}(D^{-1})^TP)\\
&=\mathbf{y}\cdot\mathbf{u}{(D^{-1})}^TP\\
&=\mathbf{u}\cdot\mathbf{y}{(D^{-1})}^T P\\
&=\mathbf{u}\cdot (D^{-1}\mathbf{y}^T)^TP\\
&=\mathbf{u}\cdot (P^T(D^{-1}\mathbf{y}^T))^T\\
&=\mathbf{u}\cdot \mathbf{y}M^{-1}
\end{align*}
Therefore from~(\ref{EQU:CM_1}), we have  
\begin{align*}
    \mathbf{y}M^{-1} \in C^{\perp}.
\end{align*}
It follows that $\mathbf{y} \in C^{\perp}M$. Hence $(CM^{-1})^{\perp} \subseteq C^{\perp}M$. Similarly, we have $C^{\perp}M \subseteq (CM^-1)^{\perp}$. Therefore, $(CM^{-1})^{\perp} = C^{\perp}M$.
\end{pf}

\noindent We now take fixed characters on~$\mathbb{F}_q$. For this, we recall~\cite{williams, gealson}.

A character~$\chi$ of~$\mathbb{F}_q$ is a homomorphism from the additive group $(\mathbb{F}_q, +)$ to the multiplicative group $(\mathbb{C}^{\ast}, \cdot)$.

Let us consider the finite field~$\mathbb{F}_q$, where $q$ is some prime power, say $q= p^m$, where $p$ is a prime number. Now suppose $f(x)$ is a primitive irreducible polynomial with degree~ $m$ over $\mathbb{F}_p$  and $f(\lambda)=0$. 
 Then we write every $\alpha \in \mathbb{F}_q$ as:
\[
\alpha = \alpha_0 + \alpha_1\lambda + \alpha_2\lambda_2 + \cdots +\alpha_{m-1}\lambda^{m-1},
\]
where $\alpha_i \in \mathbb{F}_p$ and $\chi(\alpha) :={\zeta_{p}}^{\alpha_0}$, where $\zeta_{p}$ is the primitive $p$-th root $e^{2\pi i / p}$ of unity.

The Macwilliams identities for joint weight enumerators for two~$\mathbb{F}_q$-linear codes are given as follows.
\newpage

\begin{theorem}.~\textnormal{(}\kern-0.1em\cite{williams,gealson}\textnormal{)}
\label{Theorem_MACWILLIAMS}
 Let \( C_1 \) and \( C_2 \) be two linear codes with length~$n$ over~$\mathbb{F}_q$. Also let \( C_1^{\perp} \) and \( C_2^{\perp} \) be the dual of \( C_1 \) and \( C_2 \), 
 respectively. Then we have 
 \begin{multline*}
\hspace*{-0.4cm}\mathrm{(i)}\mathbf{CJW}_{C_1^{\perp},C_2} (x_{\alpha,\beta} \mid \alpha,\beta \in \mathbb{F}_q)\\
= \frac{1}{|C_1|} \mathbf{CJW}_{C_1,C_2} \Bigg( \sum_{\mathbf{\omega} \in \mathbb{F}_q} \chi(\mathbf{\omega}\mathbf{\alpha}) x_{\omega \beta} \mid \alpha, \beta \in \mathbb{F}_q \Bigg).\hspace{3cm}
\end{multline*}
\begin{multline*}
 \hspace*{-0.4cm}\mathrm{(ii)}\mathbf{ CJW}_{C_1,C_2^{\perp}}(x_{\alpha\beta}\mid \alpha,\beta \in \mathbb{F}_q)\\
 \hspace{0.2cm}= \frac{1}{|C_2|} \mathbf{CJW}_{C_1,C_2} \Bigg( \sum_{\alpha, \beta \in \mathbb{F}_q} \chi(\beta \omega) x_{\alpha \omega}
        \mid  \alpha, \beta \in \mathbb{F}_q \Bigg).\hspace{3cm}
        \end{multline*}
        \begin{multline*}
\hspace*{-0.4cm
}\mathrm{(iii)}\mathbf{CJW}_{C_1^{\perp},C_2^{\perp}}(x_{\alpha\beta} \mid \alpha,\beta \in \mathbb{F}_q)\hspace{8.cm} \\
   \hspace{0.2cm} = \frac{1}{|C_1|}\frac{1}{|C_2|} \textbf{CJW}_{C_1,C_2} \Bigg( \sum_{\omega_1, \omega_2 \in \mathbb{F}_q}\chi(\omega_2\alpha  +  \omega_1\beta)
x_{\omega_2 \omega_1 }\mid \; \alpha, \beta \in \mathbb{F}_q \Bigg).
\end{multline*}
\end{theorem}

Now from the above Theorem~\ref{Theorem_MACWILLIAMS}, we have generalized the MacWilliams type identities for the generalized average complete joint weight enumerators of $\mathbb{F}_q$-linear codes~$C_1$ and $C_2$ as follows: 

\begin{theorem}[MacWilliams Identities]\label{Theorem:AVG_MACWILLIAMS}
 Let $C_1$ and $C_2$ be two linear codes with length~$n$ over~$\mathbb{F}_q$. Also let $C_1^{\perp}$ and $C_2^{\perp}$ be the dual of $C_1$ and $C_2$, respectively. Then we have
 \begin{multline*}
 \hspace*{-0.5cm}\mathrm{(i)} \mathbf{CJW^{av}}_{C_1^{\perp}, C_2}(x_{\alpha\beta}\mid \alpha,\beta \in \mathbb{F}_q)\\
= \frac{1}{|C_1|} \mathbf{CJW^{av}}_{C_1,C_2} \Bigg( \sum_{ \omega \in \mathbb{F}_q} \chi(\omega \alpha) x_{\omega\beta}\mid \alpha,\beta \in \mathbb{F}_q \Bigg).\hspace{3cm}
\end{multline*}
\begin{multline*}
\hspace*{-0.5cm}\mathrm{(ii)}  \mathbf{CJW^{av}}_{C_1, C_2^{\perp}}(x_{\alpha\beta}\mid \alpha,\beta \in \mathbb{F}_q)\\ 
= \frac{1}{|C_2|} \mathbf{CJW^{av}}_{C_1,C_2} \Bigg( \sum_{ \omega \in \mathbb{F}_q} \chi(\beta \omega)x_{\alpha\omega} \mid \alpha,\beta \in \mathbb{F}_q \Bigg).\hspace{3cm} 
\end{multline*}
\begin{align*}
\hspace*{-0.4cm}\mathrm{(iii)}
&\mathbf{CJW^{av}}_{C_1^{\perp}, C_2^{\perp}}(x_{\alpha\beta} \mid \alpha,\beta \in \mathbb{F}_q) \\&\hspace*{-0.1cm}= 
    \frac{1}{|C_1|} \frac{1}{|C_2|} \mathbf{CJW^{av}}_{C_1,C_2}\Bigg( 
\sum_{\omega_1, \omega_2 \in \mathbb{F}_q} \chi(\omega_2\alpha + \omega_1\beta)x_{\omega_2\omega_1} \mid  \alpha, \beta \in \mathbb{F}_q \Bigg).
\end{align*}.
\end{theorem}
\newpage
\begin{pf}
\vspace*{-0.2cm}
\begin{multline*} \text{(i)}\, \text{By definition,}\\ 
\hspace*{-5cm}\mathbf{CJW^{av}}_{C_1^{\perp}, C_2}(x_{\alpha \beta}\mid \alpha,\beta \in \mathbb{F}_q)\\
\hspace*{-0.2cm}= \frac{1}{(q-1)^n n!} \sum_{M \in \mathrm{M}_{\mathbb{F}_q}(n)} \mathbf{CJW}_{C_1^{\perp} M, C_2}(x_{\alpha \beta}\mid \alpha,\beta \in \mathbb{F}_q) \\
\hspace{0.5cm}= \frac{1}{(q-1)^n n!}\sum_{M \in \mathrm{M}_{\mathbb{F}_q}(n)}\mathbf{CJW}_{(C_1 M^{-1})^{\perp}, C_2}(x_{\alpha \beta}\mid \alpha,\beta \in \mathbb{F}_q)\\ 
\quad \hspace{7
cm}\text{[By lemma\,~\ref{Lemma:01}]} \\
\hspace*{-1.2cm}= \frac{1}{|C_1 M^{-1}|} \frac{1}{(q-1)^n n!} \sum_{M \in \mathrm{M}_{\mathbb{F}_q}(n)} \mathbf{CJW}_{C_1 M^{-1}, C_2} \\
\hspace{3.6cm}\Bigg( \sum_{\omega \in \mathbb{F}_q} \chi(\omega \alpha) x_{\omega \beta}\mid \alpha,\beta \in \mathbb{F}_q \Bigg)\\ 
\hspace*{-2.2cm}= \frac{1}{|C_1|} \frac{1}{(q-1)^n n!} \sum_{M \in \mathrm{M}_{\mathbb{F}_q}(n)} \mathbf{CJW}_{C_1 M, C_2}\\ \hspace{3.5cm}\Bigg( \sum_{\omega \in \mathbb{F}_q} \chi(\omega \alpha) x_{\omega \beta}\mid \alpha,\beta \in \mathbb{F}_q \Bigg) \\
= \frac{1}{|C_1|} \mathbf{CJW^{av}}_{C_1, C_2} \Bigg( \sum_{\omega \in \mathbb{F}_q} \chi(\omega \alpha) x_{\omega \beta}\mid \alpha,\beta \in \mathbb{F}_q \Bigg).\hspace{1.7cm}
\end{multline*}

\begin{multline*}
\text{(ii)}\, \text{By definition,}\\ 
\hspace*{-5cm}\mathbf{CJW^{av}}_{C_1, C_2^{\perp}}(x_{\alpha \beta}\mid \alpha,\beta \in \mathbb{F}_q)\\
\hspace*{-0.3cm}=\frac{1}{(q-1)^n n!} \sum_{M \in \mathrm{M}_{\mathbb{F}_q}(n)} \mathbf{CJW}_{C_1 M, C_2^{\perp}}(x_{\alpha \beta}\mid \alpha,\beta \in \mathbb{F}_q)\\
\hspace*{-2.5cm}= \frac{1}{|C_2|} \frac{1}{(q-1)^n n!} \sum_{M \in \mathrm{M}_{\mathbb{F}_q}(n)} \mathbf{CJW}_{C_1 M, C_2}\\
\hspace{3cm}\Bigg( \sum_{\omega \in \mathbb{F}_q} \chi(\beta \omega) x_{\alpha \omega}\mid \alpha,\beta \in \mathbb{F}_q
\Bigg) \\
=\frac{1}{|C_2|} \mathbf{CJW^{av}}_{C_1, C_2} \left( \sum_{\omega \in \mathbb{F}_q} \chi(\beta \omega) x_{\alpha \omega}\mid \alpha,\beta \in \mathbb{F}_q \right).\hspace{1.8cm}
\end{multline*}   
\newpage

\begin{multline*}\text{(iii)}\,\text{By definition,}\\
\hspace*{-5cm}\mathbf{CJW^{av}}_{C_1^{\perp},C_2^{\perp}}(x_{\alpha\beta} \mid \alpha,\beta \in \mathbb{F}_q) \\
\hspace*{-0.3cm}= \frac{1}{(q-1)^n n!} \sum_{M \in \mathrm{M}_{\mathbb{F}_q}(n)} 
\mathbf{CJW}_{C_1^{\perp} M,\, C_2^{\perp}}(x_{\alpha\beta} \mid \alpha,\beta \in \mathbb{F}_q) \\
\hspace{0.2cm}= \frac{1}{(q-1)^n n!} \sum_{M \in \mathrm{M}_{\mathbb{F}_q}(n)} 
\mathbf{CJW}_{(C_1 M^{-1})^{\perp},\, C_2^{\perp}}(x_{\alpha\beta} \mid \alpha,\beta \in \mathbb{F}_q)\\
\hspace{7.5cm} \text{[By Lemma~\ref{Lemma:01}]} \\
\hspace*{-1cm}= \frac{1}{(q-1)^n n!} \sum_{M \in \mathrm{M}_{\mathbb{F}_q}(n)} 
\frac{1}{|C_1 M^{-1}|}  \frac{1}{|C_2|} 
\mathbf{CJW}_{C_1 M^{-1},\, C_2}\\
\hspace{4.5cm}\left( \sum_{\omega_1,\, \omega_2 \in \mathbb{F}_q} 
\chi(\omega_2 \alpha + \omega_1 \beta) x_{\omega_2 \omega_1} 
\mid \alpha,\beta \in \mathbb{F}_q \right) \\
\hspace{7.5cm}\text{[By Theorem~\ref{Theorem_MACWILLIAMS}]} \\
\hspace*{-1.4cm}= \frac{1}{(q-1)^n n!} \sum_{M \in \mathrm{M}_{\mathbb{F}_q}(n)} 
\frac{1}{|C_1 M|} \ \frac{1}{|C_2|} 
\mathbf{CJW}_{C_1 M,\, C_2}\\
\hspace{4.5cm}\left( \sum_{\omega_1,\, \omega_2 \in \mathbb{F}_q} 
\chi(\omega_2 \alpha + \omega_1 \beta) x_{\omega_2 \omega_1} 
\mid \alpha,\beta \in \mathbb{F}_q \right) \\
\hspace*{-5.3cm}= \frac{1}{|C_1|}  \frac{1}{|C_2|} 
\mathbf{CJW^{av}}_{C_1,\, C_2} \\
\hspace{3cm}\left( \sum_{\omega_1,\, \omega_2 \in \mathbb{F}_q} 
\chi(\omega_2 \alpha + \omega_1 \beta)\, x_{\omega_2 \omega_1} 
\mid \alpha,\beta \in \mathbb{F}_q \right).
\end{multline*}

\end{pf}

\bigskip
\section{Main Result}\label{MAINTHEOREM}
In this section, we present a monomial analogue of Yoshida's theorem~\cite{Himadri} for the average complete joint weight enumerators of~$\mathbb{F}_q$-linear codes. In the beginning, we discuss about some lemmas and then we give our main theorem.  

\begin{lemma}
    Let \( C \) be a linear code with length \( n \) over \( \mathbb{F}_q \) and let \( D = \mathrm{diag}(d_1, d_2, \ldots, d_n) \) be an invertible diagonal matrix over \( \mathbb{F}_q \),
 where $d_i \in \mathbb{F}_q^{\ast}= \mathbb{F}_q\setminus \{0\}$. Then $CD$ is also a linear code over~$\mathbb{F}_q$. 
\end{lemma}
\begin{pf}
Let $C$ be a linear code with length~$n$ over~$\mathbb{F}_q$. We define 
\begin{align*}
    CD := \left\{ \mathbf{u}D = (u_1d_1, u_2d_2, \ldots, u_nd_n)\mid \mathbf{u} \in C \right\}.
\end{align*}

To show that $CD$ is a linear code over~$\mathbb{F}_q$, we show $CD$ is a vector space over $\mathbb{F}_q$. Since $C$ is a $\mathbb{F}_q$-linear code, so $C$ is an additive abelian group. It follows that, $CD$ is also an additive abelian group. Now for any $\alpha \in \mathbb{F}_q$, we have
\begin{align*}
\alpha (\textbf{u}D) &= \alpha(u_1d_1,u_2d_2,\ldots,u_nd_n) \\
&=(\alpha u_1,\alpha u_2d_2,\ldots,\alpha u_n)\mathrm{diag}(d_1,d_2,\ldots,d_n)\\
&=(\alpha\textbf{u})D \in CD.
\end{align*}
Therefore, $CD$ is a $\mathbb{F}_q$-linear code.
\end{pf}

\begin{lemma}\label{Lemma:C_D}
    Let $C$ be a linear code with length~$n$ over~$\mathbb{F}_q$ and $D=\mathrm{diag}(d_1,d_2,\ldots,d_n)$ be an invertible diagonal matrix over $\mathbb{F}_q$, where $d_i \in \mathbb{F}_q^{\ast}$. Then, 
    \begin{align*}
        \sum_{D}A_{r}^{CD} = (q-1)^{n}A_{r}^C.
    \end{align*}
\end{lemma}

\begin{pf}
    Let $C$ be a linear code with length~$n$ over~$\mathbb{F}_q$. Since $C$ and $CD$ are isomorphic to each other, therefore, it is immediate that
    \begin{align*}
        \sum_{D}A_{r}^{CD} = (q-1)^{n}A_{r}^C.
    \end{align*}
\end{pf}

\begin{theorem}[Main Theorem]\label{Theorem:Yoshida_analogous}
    Let $C_{1}$ and $C_{2}$ be two linear codes with length~$n$ over~$\mathbb{F}_q$. 
    Let~$r:= (r_{0},\ldots,r_{q-1})$ 
    and~$s := (s_{0},\ldots,s_{q-1})$ 
    be the compositions of~$n$. Also  let~$\eta$ be the bi-composition of~$n$ such that 
    \begin{align*}
        r &= \left(\sum_{\beta \in \mathbb{F}_q} \eta_{\omega_0\beta}, \ldots, \sum_{\beta \in \mathbb{F}_q} \eta_{\omega_{q-1} \beta}\right), \\
        s &= \left(\sum_{\alpha \in \mathbb{F}_q} \eta_{\alpha\omega_0}, \ldots, \sum_{\alpha \in \mathbb{F}_q} \eta_{\alpha\omega_{q-1}}\right).
    \end{align*}

    Then we have

\begin{multline*}
\mathbf{CJW^{av}}_{C_{1},C_{2}} \left(x_{\alpha\beta} \mid \alpha,\beta \in \mathbb{F}_q \right)\\
    \hspace{2cm}= \sum_{r,s,\eta} A_r^{C_{1}} A_s^{C_{2}}
     \frac{
     \prod_{i=0}^{q-1}
    \begin{array}{c}
    \left(
    \begin{array}{c}
    s_i \\
    \eta_{\omega_0 \omega_i}, \ldots, \eta_{\omega_{q-1} \omega_i}
    \end{array}
    \right)
    \end{array}
    }{
    \begin{array}{c}
    \left(
    \begin{array}{c}
    n \\
    r_0, \ldots, r_{q-1}
    \end{array}
    \right)
    \end{array}
    } \prod_{\alpha,\beta \in \mathbb{F}_q}x_{\alpha\beta}^{\eta_{\alpha\beta}},
\end{multline*}
where 
\begin{align*}
\left(
\begin{array}{c}
a \\
b_1, b_2, \ldots, b_m
\end{array}
\right) = 
\frac{
\begin{array}{c}
a!
\end{array}
}{
\begin{array}{c}
b_1! \quad b_2! \cdots b_m!
\end{array}
}.
\end{align*}
\end{theorem}

\begin{pf}
  Let $C_1$ and $C_2$ be two linear codes with length~$n$ over~$\mathbb{F}_q$. Then
   \begin{equation}\label{eq:CJWE}
    \mathbf{CJW}_{C_{1},C_{2}}(x_{\alpha\beta}\mid \alpha,\beta \in \mathbb{F}_q) 
    := 
    \sum_{\eta} 
    A_{\eta}^{C_{1},C_{2}} 
    \prod_{\alpha,\beta \in \mathbb{F}_q}
    x_{\alpha\beta}^{\eta_{\alpha\beta}}.
\end{equation}
Now we define 
\begin{equation*}
   B_{r,s,\eta}^{C_{1},\, C_{2}}
    := 
    \#\{(\mathbf{u},\mathbf{v}) \in C_{1} \times C_{2} \mid \mathrm{comp}(\mathbf{u})=r, \mathrm{comp}(\mathbf{v})=s, \eta(\mathbf{u},\mathbf{v})= \eta\}.
\end{equation*}

Hence  $A_\eta^{C_{1},C_{2}} = B_{r,s,\eta}^{C_{1},C_{2}}$,
where 
    \begin{align*}
        r &=(r_{\omega_0},r_{\omega_1},\ldots,r_{\omega_{q-1}})
        = 
        \left(
        \sum_{\beta \in \mathbb{F}_q} 
        \eta_{\omega_0\beta}, 
        \ldots, 
        \sum_{\beta \in \mathbb{F}_q} 
        \eta_{\omega_{q-1}\beta}
        \right), \end{align*}
    \begin{align*}
        s=(s_{\omega_0},s_{\omega_1},\ldots,s_{\omega_{q-1}}) 
        = 
        \left(
        \sum_{\alpha \in \mathbb{F}_q} 
        \eta_{\alpha\omega_0}, 
        \ldots, 
        \sum_{\alpha \in \mathbb{F}_q} 
        \eta_{\alpha\omega_{q-1}}
        \right).
    \end{align*}
    Therefore from ~(\ref{eq:CJWE}), we can write
 \begin{equation}
\mathbf{CJW}_{C_{1},C_{2}}(x_{\alpha\beta}\mid \alpha,\beta \in \mathbb{F}_q) 
    := 
    \sum_{r,s,\eta} 
    B_{r,s,\eta}^{C_{1},C_{2}} 
    \prod_{\alpha,\beta \in \mathbb{F}_q}
    x_{\alpha\beta}^{\eta_{\alpha\beta}}.
\end{equation} 
Now 

\begin{align*}\label{EQ:B_CM}
\hspace*{-0.5cm}\sum_{M \in \mathrm{M}_{\mathbb{F}_q}(n)} 
    B_{r,s,\eta}^{C_{1}M,\, C_{2}} 
    &= 
    \sum_{D}
    \sum_{P}
    B_{r,s,\eta}^{C_{1}DP, \, C_{2}} \nonumber\\
    & =
    \sum_{D}
    \sum_{P}
    B_{r,s,\eta}^{(C_{1}D)P,\, C_{2}} \nonumber\\
    & =
    \sum_{D}
    \#\{ (\mathbf{u},\mathbf{v}, P) \in T_r^{C_{1}D} \times T_s^{C_{2}} \times S_n \mid \eta(\mathbf{u}P, \mathbf{v}) = \eta \} \nonumber \\
    & =
    \sum_{D}
    \sum_{\substack{\mathbf{u} \in T_r^{C_{1}D} \\ \mathbf{v} \in T_s^{C_{2}}}} 
    \# \{ P \in S_n \mid \eta(\mathbf{u}P, \mathbf{v}) = \eta \} \nonumber\\
    &=(q-1)^{n}
    \sum_{\substack{\mathbf{u} \in T_r^{C_{1}} \\ \mathbf{v} \in T_s^{C_{2}}}} 
    \# \{ P \in S_n \mid \eta(\mathbf{u}P, \mathbf{v}) = \eta \}\\
     &\hspace{6cm} \text{[By Lemma~\ref{Lemma:C_D}]}
\end{align*}

Since the order of a subgroup of $S_n$ stabilizing a codeword $\mathbf{u} \in T_r^{C}$ is 
$\prod_{i = 0}^{q -1} r_i !$, then we have 
\begin{align*}
   &\sum_{M \in \mathrm{M}_{\mathbb{F}_q}(n)} 
    B_{r,s,\eta}^{C_{1}M,\, C_{2}} 
   \\& = 
    (q-1)^n
    \sum_{\substack{\mathbf{u} \in T_r^{C_{1}}, \\ \mathbf{v} \in T_s^{C_{2}}}} \prod_{i=0}^{q - 1} 
    r_i!
    \#
   \Bigg\{ 
        \mathbf{u}' \in \mathbf{F}_q^n 
        \mid 
        \textrm{comp}(\mathbf{u}') = r,\eta(\mathbf{u}', \mathbf{v}) = \eta 
    \Bigg\}\\
    &= 
    (q-1)^n
    \sum_{\mathbf{u} \in T_r^{C_{1}}} 
    \sum_{\mathbf{v} \in T_s^{C_{2}}} 
    \prod_{i=0}^{q - 1} 
    r_i!\, 
    \prod_{i=0}^{q - 1} 
    \frac{s_i!}{\prod_{j=0}^{q - 1} \eta_{\omega_j \omega_i}!}\hspace{4.45cm}
    \end{align*}
    \begin{align*}
     &=
    (q-1)^n 
    A_r^{C_{1}} 
A_s^{C_{2}}\, 
    n!\,    \frac{\prod_{i=0}^{q-1} \dfrac{s_i!}{\prod_{j=0}^{q-1} \eta_{\omega_j\omega_i}!}}{\dfrac{n!}{\prod_{i=0}^{q-1} r_i!}}\\
    &=
    (q-1)^n\, n!\,
    A_r^{C_{1}} 
    A_s^{C_{2}} 
    \frac{{\prod_{i=0}^{q -1}}
    \begin{array}{c}
    \left(
    \begin{array}{c}
    s_i \\
    \eta_{\omega_0 \omega_i}, \ldots, \eta_{\omega_{q-1} \omega_i}
    \end{array}
    \right)
    \end{array}
    }{
    \begin{array}{c}
    \left(
    \begin{array}{c}
    n \\
    r_0, \ldots, r_{q-1}
    \end{array}
    \right)
    \end{array}.
        }.\hspace{3.5cm}
\end{align*}
Therefore
\begin{align*}
&\mathbf{CJW^{av}}_{C_{1},C_{2}}(x_{\alpha\beta} \mid \alpha,\beta \in \mathbb{F}_q) \\
&= 
    \frac{1}{(q-1)^n n!} 
    \sum_{M \in \mathrm{M}_{\mathbb{F}_q}(n)} 
    \mathbf{CJW}_{C_{1}M, C_{2}}(x_{\alpha\beta}\mid \alpha,\beta \in \mathbb{F}_q)\hspace{2.8cm} \\
    &= 
    \frac{1}{(q-1)^n n!}
    \sum_{r,s,\eta}
    B_{r,s,\eta}^{C_{1}M,\, C_{2}} 
    \prod_{\alpha,\beta \in \mathbb{F}_q} x_{\alpha\beta}^{\eta_{\alpha\beta}} \\
    &=
    \sum_{r,s,\eta}
    A_r^{C_{1}} 
    A_s^{C_{2}} 
    \frac{\prod_{i=0}^{q-1}
    \left(
    \begin{array}{c}
    s_i \\
    \eta_{\omega_0 \omega_i}, \ldots, \eta_{\omega_{q-1} \omega_i}
    \end{array}
    \right)}
    {
    \left(
    \begin{array}{c}
    n \\
    r_0, \ldots, r_{q-1}
    \end{array}
    \right)
    }
    \prod_{\alpha,\beta \in \mathbb{F}_q} x_{\alpha\beta}^{\eta_{\alpha\beta}}.
\end{align*}
Hence the proof is completed.
 \end{pf}
 
\section[Average $g$-fold complete joint weight enumerators]{\texorpdfstring{Average $g$-fold\\ complete joint weight enumerators}{Average $g$-fold complete joint weight enumerators}}\label{Section:G-fold}

 In this section, we present a generalized version of main theorem for average $g$-fold complete joint weight enumerators for $\mathbb{F}_q$-linear codes.

 Let $C_1, C_2,\ldots, C_g$ be linear codes of length $n$ over $\mathbb{F}_q$. For $g$-tuple $(\mathbf{c}_1, \mathbf{c}_2,\ldots, \mathbf{c}_g) \in C_1\times C_2\times \cdots \times C_g$, we define the $g$-fold composition $\eta_(\mathbf{c}_1,\mathbf{c}_2,\ldots, \mathbf{c}_g)$ as 
 \begin{align*}
    \eta_(\mathbf{c}_1,\mathbf{c}_2,\ldots, \mathbf{c}_g)^{g} :=( \eta_{a}^{g}(\mathbf{c}_1,\mathbf{c}_2,\ldots, \mathbf{c}_g)\mid a \in \mathbb{F}_q^{g}), 
 \end{align*}
where each $\eta_{a}^{g}(\mathbf{c}_1,\mathbf{c}_2,\ldots, \mathbf{c}_g)$ is a non-negative integers and 
\begin{align*}
    \eta_{a}^{g}(\mathbf{c}_1,\mathbf{c}_2,\ldots, \mathbf{c}_g) := \# \{ i \mid (c_{1_i}, c_{2_i},\ldots, c_{g_i}) = a\}
\end{align*}
Clearly, $\sum_{a\in \mathbb{F}_q^g}\eta_{a}^g = n$. Also we define 
\begin{align*}
    T_{\eta_g}^{C_1,C_2.\ldots, C_g} := \#\{ (\mathbf{c}_1,\mathbf{c}_2,\ldots, \mathbf{c}_g) \in C_1 \times C_2\times \cdots \times C_g\\ \mid \eta_{g}(\mathbf{c}_1,\mathbf{c}_2,\ldots, \mathbf{c}_g) = \eta_{g}\}
\end{align*}
The $g$-fold complete joint weight enumerator is defined as 
\begin{align*}
    \mathbf{CJW}_{C_1,C_2\ldots,C_g}(x_a \mid a \in \mathbb{F}_q^g) &:= \sum_{\mathbf{c}_1\in C_1,\ldots, \mathbf{c}_g\in C_g}\prod_{a \in \mathbb{F}_q^g}x_{a}^{\eta_{a}^{g}(\mathbf{c}_1,\mathbf{c}_2,\ldots, \mathbf{c}_g)}\\
    &= \sum_{\eta_g}A_{\eta_g}^{C_1,C_2,\ldots,C_g} \prod_{a \in \mathbb{F}_q^g}x_{a}^{\eta_{a}^g},
\end{align*}
where $x_a$ are indeterminates and ~$A_{\eta_g}^{C_1,C_2,\ldots,C_g}:= \# \{(\mathbf{c}_1,\mathbf{c}_2,\ldots, \mathbf{c}_g \in C_1\times C_2\times \cdots \times C_g \mid \eta(\mathbf{c}_1,\mathbf{c}_2\ldots,\mathbf{c}_g)^g = \eta^{g}$\}. The average~$g$-fold complete joint weight enumerator is defined as 
\begin{multline*}
    \mathbf{CJW}^{\mathrm{av}}_{C_1,C_2,\ldots,C_g}(x_a \mid a \in \mathbb{F}_q^g)
    \\:= \frac{1}{(q-1)^n n!} \sum_{M \in M_{\mathbb{F}_q}(n)} \mathbf{CJW}_{C_1M,C_2,\ldots,C_g}(x_a \mid a \in \mathbb{F}_q^g)
\end{multline*}

For $a=(a_1,a_2,\ldots,a_g)\in \mathbb{F}_q^g$ and $b=(b_1,b_2,\ldots,b_{g-1}) \in \mathbb{F}_q^{g-1}$, we denote 
\begin{align*}
    &[a;j]:= (a_1,a_2,\ldots,a_{j-1},a_{j+1}, \ldots, a_{g}) \in \mathbb{F}_q^{g},\\
    &(z,b):= (z, b_1,b_2, \ldots, b_{g-1}) \in \mathbb{F}_q^{g}\quad \text{for}\, z \in \mathbb{F}_q
\end{align*}
Now we present the generalization of Theorem~\ref{Theorem:Yoshida_analogous}.
\begin{theorem}\label{Theorem:g-fold_Yoshida}
    Let $C_1,C_2,\ldots, C_g$ be $g$ linear codes of length~$n$ over~$\mathbb{F}_q$. Also let $s_1, s_2,\ldots,s_g$ be the composition of $n$. Consider $\eta^{g}$ be the~$g$-fold composition of $n$ such that for $j=1,2,\ldots,g$ we have 
    \begin{align*}
    s_{j} = \left(\sum_{a \in \mathbb{F}_q^g}\eta^{g}_{a}\mid a_j=\omega_i \quad \text{for } i = 0, 1, \ldots, q - 1 \right) 
\end{align*}
and $\eta^{g-1}$ is the~$(g-1)$-fold composition of $n$ such that 
\begin{align*}
    \eta^{g-1}:= (\eta^{g-1}_{b} \quad \text{for}\,\, b \in \mathbb{F}_q^{g-1}),
\end{align*}
where each $\eta^{g-1}_{b}$ is  non-negative integers and $\eta^{g-1}_{b}= \sum_{a\in \mathbb{F}_q^g}\eta_{a_{[a;1]=b}}^g$. Then we have 
\begin{multline*}
    \mathbf{CJW}^{av}(x_a \mid a \in \mathbb{F}_q^g) \\
    = \sum_{s_1,\eta^{g-1},\eta^g} A_{s_1}^{C_1}A_{\eta^{g-1}}^{C_2,\ldots,C_g}   \frac{
     \prod_{b \in \mathbb{F}_q^g}
    \begin{array}{c}
    \left(
    \begin{array}{c}
    \eta_{b}^{g-1} \\
    \eta^{g}_{(\omega_0,b)}, \ldots, \eta^{g}_{(\omega_{q-1},b)}
    \end{array}
    \right)
    \end{array}
    }{
    \begin{array}{c}
    \left(
    \begin{array}{c}
    n \\
    s_{10}, \ldots, s_{1q-1}
    \end{array}
    \right)
    \end{array}
    }\prod_{a \in \mathbb{F}_q^g} x_{a}^{\eta^{g}_{a}}
\end{multline*}
where 
\begin{align*}
\left(
\begin{array}{c}
p \\
r_1, r_2, \ldots, r_m
\end{array}
\right) = 
\frac{
\begin{array}{c}
p!
\end{array}
}{
\begin{array}{c}
r_1! \quad r_2! \cdots r_m!
\end{array}
}.
\end{align*}

    \end{theorem}
    \begin{pf}
        Let $C_1,C_2,\ldots, C_g$ be the $g$ linear codes of length~$n$ over~$\mathbb{F}_q$. Then by the definition of $g$-fold complete joint weight enumerators of $C_1, C_2, \ldots, C_g$, we have 
\begin{equation}\label{EQ:g-fold}
\mathbf{CJW}_{C_1,C_2, \ldots, C_g}(x_a\mid a \in \mathbb{F}_q^g):=  
    \sum_{\eta_g}A_{\eta_g}^{C_1,C_2,\ldots,C_g} 
    \prod_{a \in \mathbb{F}_q^g}x_{a}^{\eta_{a}^g},
\end{equation}

       where $\sum_{a\in \mathbb{F}_q^g}\eta_{a}^g= n$. Now we define 
        \begin{align*}
            B_{s_1,\eta_{g-1}\eta_g}^{C_1,C_2,\ldots,C_g} := \#\{ (\mathbf{c}_1,\mathbf{c}_2\ldots, \mathbf{c}_g) \in C_1\times C_2 \cdots \times C_g \mid comp(\mathbf{c_1})= s_1, \eta^{g-1}\\(\mathbf{c}_2,\ldots,\mathbf{c}_g )= \eta^{g-1}, \eta^{g}(\mathbf{c}_1, \ldots, \mathbf{c}_g) = \eta^{g}\}.
        \end{align*}
        Therefore, 
        \begin{align*}
            A_{\eta_{g}}^{C_1,C_2,\ldots,C_g} = B_{s_1,\eta^{g-1},\eta^{g}}^{C_1,C_2,\ldots,C_g}.
        \end{align*}
        Hence from~(\ref{EQ:g-fold}), we can write 
        \begin{align*}
            \mathbf{CJW}_{C_1,C_2,\ldots,C_g}(x_a \mid a \in \mathbb{F}_q^g) =\sum_{s_1, \eta^{g-1},\eta^{g}}B_{s_1,\eta^{g-1},\eta^{g}}^{C_1,C_2,\ldots,C_g}\prod_{a \in \mathbb{F}_q^g}x_a^{\eta^{g}_{a}}. 
        \end{align*}
        Now 
        \begin{multline*}
    \sum_{M \in M_{\mathbb{F}_q(n)}} B_{s_1, \eta^{g-1}, \eta^{g}}^{C_1,C_2,\ldots, C_g} \\
    \hspace*{-5cm} = \sum_{D} \sum_{P} B_{s_1, \eta_{g-1}, \eta_{g}}^{C_1 D P, C_2, \ldots, C_g} \\
    \hspace*{-5cm} = \sum_{D} \sum_{P} B_{s_1, \eta_{g-1}, \eta_{g}}^{(C_1 D) P, C_2, \ldots, C_g} \\
    \hspace{1cm} = \sum_{D} \#\left\{ (\mathbf{c}_1, \mathbf{c}_2, \ldots, \mathbf{c}_g, P) \in T_{s_1}^{C_1} \times T_{s_2}^{C_2} \times \cdots \times T_{s_g}^{C_g} \times S_n \right. \\
    \left. \hspace{5cm} \mid \eta^{g}(\mathbf{c}_1, \mathbf{c}_2, \ldots, \mathbf{c}_g ) = \eta^{g} \right\} \\
    \hspace*{-4cm} = \sum_{D} \sum_{\mathbf{c}_1 \in T_{s_1}^{C_1}} \sum_{(\mathbf{c}_2, \mathbf{c}_3, \ldots, \mathbf{c}_g) \in T_{\eta_{g-1}}^{C_2,\ldots,C_g}}\\ \hspace{2cm}\#\{P \in S_n\mid \eta^{g}(\mathbf{c}_1P, \ldots, \mathbf{c}_g) = \eta^{g}\}\\
    \hspace*{-3cm}=(q-1)^n \sum_{\mathbf{c}_1 \in T_{s_1}^{C_1}} \sum_{(\mathbf{c}_2, \mathbf{c}_3, \ldots, \mathbf{c}_g) \in T_{\eta_{g-1}}^{C_2,\ldots,C_g}} \\ \#\{P \in S_n\mid \eta^{g}(\mathbf{c}_1P, \ldots, \mathbf{c}_g) = \eta^{g}\}\\
    \hspace{6cm} \text{[By Lemma~\ref{Lemma:C_D}]}
\end{multline*}
Since the order of a subgroup of $S_n$ stabilizing $\mathbf{c}_1 \in T_{s_1}^{C_1}$ is $\prod_{i=0}^{q-1} s_{1_i}$, then we have
\begin{multline*}
    \sum_{M \in M_{\mathbb{F}_q}(n)} B_{s_1, \eta^{g-1}, \eta^{g}}^{C_1,C_2,\ldots,C_g} \\
    \hspace*{-3.5cm}= (q-1)^n \sum_{\mathbf{c}_1 \in T_{s_1}^{C_1}} \sum_{(\mathbf{c}_2, \ldots, \mathbf{c}_g) \in T_{\eta_{g-1}}^{C_2,\ldots,C_g}} 
    \prod_{i=0}^{q-1} s_{1,i} \\
    \hspace{2.5cm} \#\left\{ \mathbf{c}_1' \in \mathbb{F}_q^n \,\middle|\, \operatorname{comp}(\mathbf{c}_1') = s_1,\ 
    \eta^g(\mathbf{c}_1', \mathbf{c}_2, \ldots, \mathbf{c}_g) = \eta^g \right\} \\
    =(q-1)^n n! A_{s_1}^{C_1}A_{\eta_{g-1}}^{C_2,C_3,\ldots,C_g} \prod_{i=0}^{q-1} s_{1_i}\frac{
     \prod_{b \in \mathbb{F}_q^g}
    \begin{array}{c}
    \left(
    \begin{array}{c}
    \eta_{b}^{g-1} \\
    \eta^{g}_{(\omega_0,b)}, \ldots, \eta^{g}_{(\omega_{q-1},b)}
    \end{array}
    \right)
    \end{array}
    }{
    \begin{array}{c}
    \left(
    \begin{array}{c}
    n \\
    s_{10}, \ldots, s_{1q-1}
    \end{array}
    \right)
    \end{array}
    }
\end{multline*}
Therefore, we have 
\begin{multline*}
    \mathbf{CJW}^{av}_{C_,1,C_2,\ldots,C_g}(x_a \mid a \in \mathbb{F}_q^n) \\
    \hspace*{-1.5cm}= \frac{1}{(q-1)^n n!} \sum_{M \in M_{\mathbb{F}_q}(n)} \mathbf{CJW}_{C_1M, C_2, \ldots, C_g}(x_a\mid a \in \mathbb{F}_q^n) \\
   \hspace*{-3.3cm} = \frac{1}{(q-1)^n n!} \sum_{s_1, \eta^{g-1}, \eta^g} B_{s_1, \eta^{g-1}, \eta^g}^{C_1M, C_2, \ldots, C_g} \prod_{a \in \mathbb{F}_q^n}x_{a}^{\eta_{a}^g}\\
    = \sum_{s_1, \eta^{g-1}, \eta_g} A_{S_1}^{C_1}A_{\eta_{g-1}}^{C_2, \ldots, C_g} \frac{
     \prod_{b \in \mathbb{F}_q^g}
    \begin{array}{c}
    \left(
    \begin{array}{c}
    \eta_{b}^{g-1} \\
    \eta^{g}_{(\omega_0,b)}, \ldots, \eta^{g}_{(\omega_{q-1},b)}
    \end{array}
    \right)
    \end{array}
    }{
    \begin{array}{c}
    \left(
    \begin{array}{c}
    n \\
    s_{10}, \ldots, s_{1q-1}
    \end{array}
    \right)
    \end{array}
    }\prod_{a \in \mathbb{F}_q^g} x_{a}^{\eta^{g}_{a}}
\end{multline*}
This completes the proof.
\end{pf}

\section*{Acknowledgments}
The author thanks Dr Himadri Shekhar Chakraborty
for helpful discussions.

\end{document}